\documentclass[titlepage,11pt]{article}

\pdfoutput=1

\usepackage{amssymb,amsmath,amsfonts}
\usepackage{bm}
\usepackage{epsfig}
\usepackage{ccaption}
\usepackage{graphicx}
\usepackage{float}
\usepackage{breqn}

\usepackage[
      colorlinks=true,
      linkcolor=blue,
      urlcolor=blue,
      filecolor=black,
      citecolor=red,
      pdfstartview=FitV,
      pdftitle={},
        pdfauthor={Roberto Emparan},
        pdfsubject={},
        pdfkeywords={},
        pdfpagemode=None,
        bookmarksopen=true
      ]{hyperref}

\newcommand{\beq}{\begin{equation}}
\newcommand{\eeq}{\end{equation}}
\newcommand{\beqa}{\begin{eqnarray}}
\newcommand{\eeqa}{\end{eqnarray}}
\newcommand{\bea}{\begin{eqnarray}}
\newcommand{\eea}{\end{eqnarray}}

\newcommand{\nn}{\nonumber}

\newcommand{\ie}{{\it i.e.,\,}}
\newcommand{\eg}{{\it e.g.,\,}}

\newcommand{\lp}{\left(}
\newcommand{\rp}{\right)}

\newcommand{\ord}[1]{{\mathcal O}\lp #1\rp}

\newcommand{\sR}{\mathsf{R}}

\newcommand{\hr}{\hat{r}}
\newcommand{\htt}{\hat{t}}

\newcommand{\hph}{\hat{\phi}}
\newcommand{\br}{\rho}

\newcommand{\nh}{m_\phi}

\newcommand{\fq}{\mathfrak{q}}
\newcommand{\aq}{a_q}

\newcommand{\cR}{\mathcal{R}}

\numberwithin{equation}{section}

\def\clock{{\count0=\time
           \divide\count0 60
           \ifnum\count0<10 0\fi\the\count0
           \multiply\count0 -60 \advance\count0 \time
           :\ifnum\count0<10 0\fi \the\count0
         }}
\newcommand{\timestamp}{{\small\vbox{\hbox{\tt \jobname.pdf}
\hbox{\the\day/\the\month/\the\year, \clock
}}}}

\begin{document}

\begin{titlepage}
\leftline{}
\vskip 2cm
\centerline{\LARGE \bf Charged rotating black holes in higher dimensions}
\vskip 1.2cm
\centerline{\bf Tom\'as Andrade$^{a}$, Roberto Emparan$^{a,b}$, David Licht$^{a}$}
\vskip 0.5cm
\centerline{\sl $^{a}$Departament de F{\'\i}sica Qu\`antica i Astrof\'{\i}sica, Institut de
Ci\`encies del Cosmos,}
\centerline{\sl  Universitat de
Barcelona, Mart\'{\i} i Franqu\`es 1, E-08028 Barcelona, Spain}
\smallskip
\centerline{\sl $^{b}$Instituci\'o Catalana de Recerca i Estudis
Avan\c cats (ICREA)}
\centerline{\sl Passeig Llu\'{\i}s Companys 23, E-08010 Barcelona, Spain}
\smallskip
\vskip 0.5cm
\centerline{\small\tt tandrade@icc.ub.edu, \, emparan@ub.edu,\, david.licht@icc.ub.edu}

\vskip 1.2cm
\centerline{\bf Abstract} \vskip 0.2cm 
\noindent 
We use a recent implementation of the large $D$ expansion in order to construct the higher-dimensional Kerr-Newman black hole and also new charged rotating black bar solutions of the Einstein-Maxwell theory, all with rotation along a single plane. We describe the space of solutions, obtain their quasinormal modes, and study the appearance of instabilities as the horizons spread along the plane of rotation. Generically, the presence of charge makes the solutions less stable. Instabilities can appear even when the angular momentum of the black hole is small, as long as the charge is sufficiently large. We expect that, although our study is performed in the limit $D\to\infty$, the results provide a good approximation for charged rotating black holes at finite $D\geq 6$.

\end{titlepage}
\pagestyle{empty}
\small
\normalsize
\newpage
\pagestyle{plain}
\setcounter{page}{1}

%



\section{Introduction}

Rotation and charge often have similar effects on a black hole, both of them opposing the gravitational field attraction. The Kerr-Newman solution indeed shows that, for a given mass, the black hole reduces its size as either charge or rotation are added \cite{Newman:1965my}. 

In higher dimension $D\geq 5$ one expects that these behaviors not only persist but become more varied, since the gravitational effects of rotation and charge have different fall-off with distance. It is natural to anticipate a rich spectrum of black hole physics as charge and rotation are increased, extending what is already known when only rotation is present \cite{Emparan:2008eg}.
However, the investigation of this problem has been hampered by the striking fact that, long after Tangherlini extended the Reissner-Nordstr{\"o}m solution to any $D\geq 5$ \cite{Tangherlini:1963bw}, and Myers and Perry did likewise for the Kerr solution \cite{Myers:1986un}, the only charged and rotating black hole solution of the Einstein-Maxwell equations known exactly in any $D\geq 4$ remains the Kerr-Newman solution\footnote{In this article we only consider asymptotically flat black hole solutions of the pure Einstein-Maxwell theory, without any Chern-Simons terms nor additional scalar fields.}. 

Although approximate solutions have been obtained through a variety of methods\footnote{These include: perturbatively small charge \cite{NavarroLerida:2007ez,Allahverdizadeh:2010xx,Allahverdizadeh:2010fn,Ortaggio:2006ng} or slow rotation \cite{Horne:1992zy,Aliev:2005npa,Aliev:2006yk,Chen:2017wpf}; charged rotating black holes with two widely separate horizon length scales \cite{Caldarelli:2010xz,Emparan:2011hg,Armas:2013aka}; charged black holes in large odd $D$ \cite{Tanabe:2016opw} or at large $D$ with either small charge or rotation \cite{Mandlik:2018wnw,Kundu:2018dvx}.}, large regions in parameter space remain poorly explored where unusual features of black holes may be revealed. We find particularly appealing the discovery in \cite{Caldarelli:2010xz} of near-extremal charged black holes in all $D\geq 6$ which have arbitrarily small spin but are nevertheless very far from the extremal Reissner-Nordstr{\"o}m black hole. Their horizons are not approximately round but are instead highly pancaked along the rotation plane, even though their angular momentum and angular velocity are small. In the extremal limit they approach a static, singular disk of charged dust. Little else is known about these black holes. The methods of \cite{Caldarelli:2010xz} only yield access to certain limits of parameter space and do not allow, \eg to connect between the Reissner-Nordstr{\"o}m solution and these pancaked near-extremal black holes. Interestingly, although the latter are arbitrarily close to extremality, they were conjectured in \cite{Caldarelli:2010xz} to be dynamically unstable due to an undamped quasinormal mode. Unfortunately, the quasinormal spectra of black holes with both charge and rotation are almost unknown\footnote{All the studies that we know of use the large $D$ expansion: \cite{Chen:2017wpf,Kundu:2018dvx} for small charge or rotation, and \cite{Tanabe:2016opw} for black holes with all rotations turned on in odd $D$.}. 

In this article we show that many of these limitations can be overcome through the use of the large-$D$ effective theory of black holes \cite{Asnin:2007rw,Emparan:2013moa,Emparan:2014aba,Emparan:2015hwa,Emparan:2016sjk,Bhattacharyya:2015dva,Bhattacharyya:2015fdk}, and more specifically its recent efficient implementation in \cite{Andrade:2018nsz}. As we will see, these methods allow to analytically investigate black hole phases with arbitrary values of the charge and of the angular momentum (in a single plane), as well as to obtain their quasinormal spectrum at low frequencies, $\omega =\ord{D^0}$. Rotations in any finite number of planes are straightforward to add, but we will not pursue this in this article.

We expect that our results are not only qualitatively but also quantitatively good in any $D\geq 6$. This is a reasonable prospect given the remarkable accuracy that the large-$D$ methods have obtained for the features, including instability onsets and unstable growth rates, of neutral rotating black holes in this range of dimensions \cite{Andrade:2018nsz,Suzuki:2015iha}. In contrast, rotating black holes in $D=5$ typically exhibit qualitative differences compared to $D=4$ and $D\geq 6$; it would not be surprising if the non-perturbative corrections to the $1/D$ expansion became large in $D=5$.

The approach started in \cite{Andrade:2018nsz} for the investigation of black holes succeeds by focusing on the region of the horizon where, when $D$ is large, most of the physics of the black hole concentrates: a small cap of polar-angular extent $\Delta\theta=\ord{1/\sqrt{D}}$ around the rotation axis. Here the horizon is well approximated by a gaussian bulge on a black membrane. That is, we study lumps on a black brane that share the main properties of a localized black hole. The study of this bulge accurately reproduces properties of the black hole such as its shape, area, mass, and angular momentum. Moreover, its linear perturbations can be solved to obtain the waveforms and frequencies of the least-damped quasinormal modes. In addition, this approach has revealed the existence of `black bar' configurations: elongated, bar-shaped rotating black holes whose emission of gravitational radiation vanishes to all perturbative orders in $1/D$ and therefore evolve very slowly at finite but large $D$.  

We will see that the methods of \cite{Andrade:2018nsz} readily extend to black holes with charge in addition to rotation ---surprisingly easily, given the difficulties in doing so at any finite $D$. Although technically the extension is straightforward, the resulting parameter space becomes richer with the inclusion of charge. In particular, we will be able to interpolate between the static, spherical Reissner-Nordstr{\"o}m black hole, and the near-extremally charged, small-spin pancaked black holes of \cite{Caldarelli:2010xz}. Along the way, we will identify the quasinormal modes that trigger the instability of the latter. We will also construct a new family of charged black bars, which remain stationary since their radiation into photons and gravitons is exponentially suppressed in the $1/D$ expansion.

\begin{figure}[th]
	\centerline{\includegraphics[width=1\textwidth]{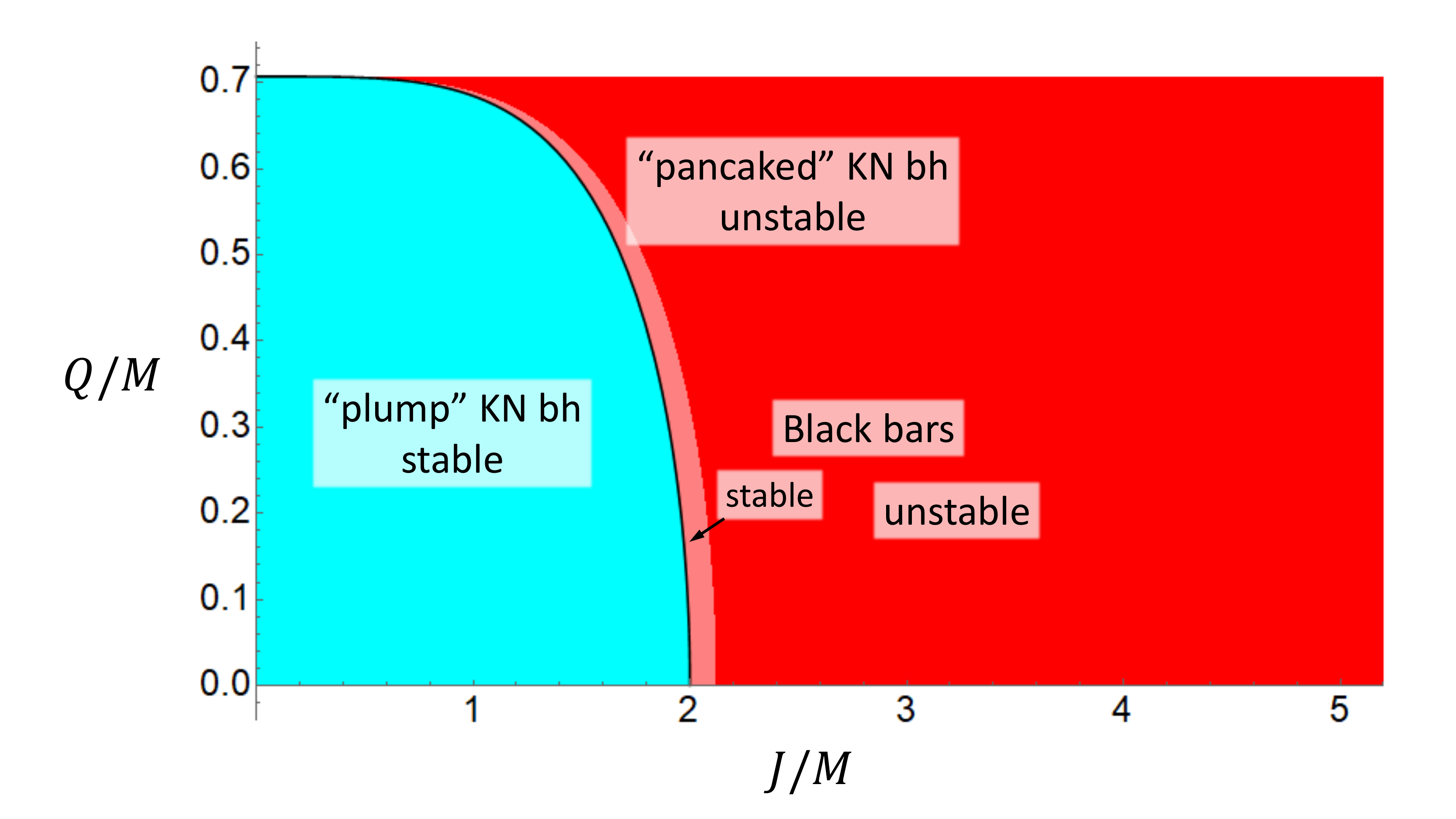}}
	\caption{\small Phases of the charged rotating black holes constructed in this article to leading order in the $1/D$ expansion. We label them by their angular momentum $J$ and charge $Q$ for fixed mass $M$. The charge is bounded above, $Q\leq M/\sqrt{2}$, but the upper, extremal limit lies outside the strict range of validity of our construction. Kerr-Newman-like (KN) black holes exist for all $J/M$. The black line $J/M=2\lp 1-2(Q/M)^2\rp^{1/4}$ separates the region (blue) where they are stable and round-shaped (``plump''), from the region (red, light and dark) where they are unstable and pancake-shaped (eq.~\eqref{JMzero}). Charged black bars exist in the red regions (sec.~\ref{sec:nuniq}), but are stable (and only to leading order in $1/D$) only in the light-red area with outer boundary $J/M=(3/\sqrt{2})\lp 1-2(Q/M)^2\rp^{1/4}$ (eq.~\eqref{JMzerobar}). In the upper-left corner there exist near-extremal black holes with arbitrarily small spin (sec.~\ref{sec:effQJ}): below the black line they are close to the Reissner-Nordstr{\"o}m solution, approximately spherical and stable; above the black line they are unstable and highly pancaked.\label{fig:Phases}}
\end{figure}

Figure~\ref{fig:Phases} summarizes our main findings concerning the space of solutions of charged rotating black holes. 
We refer to the axisymmetric charged rotating solutions as ``the large-$D$ limit of the higher-dimensional Kerr-Newman black hole''. This is justified since these solutions correctly reproduce the limit $D\to\infty$ of the Reissner-Nordstr{\"o}m-Tangherlini black hole when $J=0$, and of the Myers-Perry black hole when $Q=0$.

Our plan is as follows. In the next section we present the formalism that we employ throughout the article. In sec.~\ref{sec:KN} we construct and study the solutions that correspond to the limit $D\to\infty$ of the Kerr-Newman black hole. Sec.~\ref{sec:bbars} describes charged black bars. Sec.~\ref{sec:qnms} computes the quasinormal modes and stability properties of the solutions of the previous sections. We then conclude in sec.~\ref{sec:outlook}. In appendix~\ref{app:RNT} we show that the Reissner-Nordstr{\"o}m-Tangherlini solution is correctly recovered with our methods in the limit $D\to\infty$.

\section{Effective equations}

We study electrically charged black holes of the Einstein-Maxwell theory
\beq\label{chaction}
I=\int d^D x\sqrt{-g}\lp R-\frac14 F^2\rp \,,
\eeq
in the limit of large $D$.
Setting
\beq
D=n+p+3\,,
\eeq
with finite $p$, and expanding in series of $1/n$, ref.~\cite{Emparan:2016sjk} investigated solutions in the form of fluctuating black $p$-branes extended along spatial directions $\sigma^i$, $i=1,\dots,p$ ,
\beq\label{AFch}
ds^2=2dtdr-Adt^2-\frac{2}{n} C_i d\sigma^idt+\frac1{n}G_{ij}d\sigma^id\sigma^j +r^2d\Omega_{n+1}\,,
\eeq
where $\sR=r^n$. The metric coefficients are
\beq
A=1-\frac{m(t,\sigma)}{\sR}+\frac{q(t,\sigma)^2}{2\sR^2}\,,\qquad 
C_i=\left(1-\frac{q(t,\sigma)^2}{2m(t,\sigma)\sR}\right) \frac{p_i(t,\sigma)}{\sR}\,,
\eeq 
\begin{align}
G_{ij}&=\delta_{ij}+\frac1{n}\left\{\left( 1-\frac{q(t,\sigma)^2 }{2m(t,\sigma)\sR}\right)\frac{p_i(t,\sigma) p_j(t,\sigma)}{m(t,\sigma)\sR} 
\right. \nonumber \\
& \hphantom{{} = \delta_{ij}+\frac1{n}\,}
\left.
-\ln\lp 1-\frac{m_-(t,\sigma)}{\sR}\rp
\left[ 2\delta_{ij}+  \nabla_i \frac{p_j(t,\sigma)}{m(t,\sigma)} +\nabla_j \frac{p_i(t,\sigma)}{m(t,\sigma)}\right] 
 \right\} \,,
\end{align}
and the electric potential is
\beq
A_t = -\frac{q(t,\sigma)}{\sR}\,.
\eeq

The worldvolume collective fields $m(t,\sigma)$, $q(t,\sigma)$ are the mass and charge density of the black brane. The fields $p_i(t,\sigma)$ are often conveniently traded for velocities $v_i(t,\sigma)$ through
\beq
p_i = m v_i+\nabla_i m\,.
\eeq
We also find it convenient to define
\beq
m_\pm =\frac{1}{2}\lp m\pm \sqrt{m^2-2q^2}\rp\,.
\eeq

The Einstein-Maxwell equations are solved to $\ord{1/D}$ if and only if these collective fields satisfy the effective equations for mass continuity,
\beq
\label{mass continuity}
\partial_t m+\nabla_i (m v^i)=0\,,
\eeq
for momentum continuity,
\beq
\partial_t (m v^i)+\nabla_j (m v^i v^j +\tau^{ij})=0
\eeq
with stress tensor
\beq\label{chstress}
\tau_{ij}= -\lp m_+ -m_- \rp\delta_{ij}  -2m_+\nabla_{(i}v_{j)}-(m_+-m_-)\,\nabla_i\nabla_j \ln m \,,
\eeq
and for charge continuity,
\beq
\label{dtq}
\partial_t q+\nabla_i j^i=0
\eeq
with current
\beq\label{chj}
j_i=q v_i -m\nabla_i\lp\frac{q}{m}\rp\,.
\eeq
Note that the one-derivative terms in $\tau_{ij}$ and $j_i$ can be interpreted as viscous stresses and charge diffusivities. It has been proven in \cite{Emparan:2016sjk} that, with these equations, charge diffusion leads to entropy production but viscosity does not. 

The solutions have an outer event horizon at $\sR=m_+(t,\sigma)$ wherever and whenever $m(t,\sigma)>\sqrt{2}q(t,\sigma)$. In principle the extremal limit $m= \sqrt{2}q$ lies outside the range of validity of the approximations made in the derivation. Taking the extremal limit requires separate study, and therefore in this article we will always remain strictly away from it.

\subsection{Stationary configurations}\label{subsec:statconf}

Following \cite{Andrade:2018nsz}, we investigate stationary configurations where the mass and charge density are Lie-dragged with velocity $v^i$,
\beq
(\partial_t+v^i\partial_i) m=0\,,\qquad (\partial_t+v^i\partial_i) q=0\,,
\eeq
but without acceleration, $\partial_t v^i=0$. In addition we require that dissipative effects are absent, be they viscous shear and expansion or charge diffusion, so that
\beq
\nabla_{(i}v_{j)}=0
\eeq
and
\beq
\nabla_i\lp\frac{q}{m}\rp=0\,.
\eeq
The latter implies that
\beq\label{gothq}
\fq \equiv \frac{q}{m}
\eeq
is a constant, so that the charge density must proportionally track the mass density exactly. This property of the charge distribution in stationary configurations provides the crucial simplification that will allow us to easily obtain charged rotating black hole solutions from neutral ones.

In order to see how this occurs, we derive a single master equation for stationary configurations in terms of the area-radius variable
\beq
\cR=\ln m\,.
\eeq
The derivation piggybacks on \cite{Emparan:2016sjk} and \cite{Andrade:2018nsz} to arrive at
\beq
\nabla_i\lp \frac{v^2}{2}+\frac{m_+-m_-}{m}\lp\mathcal{R}+\nabla_j\nabla^j\mathcal{R}+\frac{1}{2}\nabla^j\mathcal{R}\nabla_j\mathcal{R}\rp\rp =0\,.
\eeq
Noting that \eqref{gothq} implies that $m_\pm/m$ are constants, we define a `charge-rescaled velocity'
\beq\label{vq}
v^i_q=\sqrt{\frac{m}{m_+-m_-}}\,v^i=\frac{v^i}{\lp 1-2\fq^2\rp^{1/4}}\,.
\eeq
Then, after absorbing an integration constant by shifting $\cR$ (which is simply a rescaling of the mass), the master equation takes the form
\begin{equation}\label{mastercharge}
\frac{v_q^2}{2}+\mathcal{R}+\nabla_j\nabla^j\mathcal{R}+\frac{1}{2}\nabla^j\mathcal{R}\nabla_j\mathcal{R}=0\,.
\end{equation}

All of the dependence on the charge in this equation is encoded in $v_q$. Therefore, given a neutral stationary solution for $\cR$ with velocity $v$, we can immediately construct a charged stationary solution by substituting $v\to v_q$. Note that this substitution must not be applied when $v$ appears through the comoving dependence on $\sigma^i -v^i t$, since this is fixed by the stationarity condition of invariance under $\partial_t+ v^i\partial_i$.

After obtaining in this manner the mass density $m=\exp \cR$ for the new charged solution in terms of $v_q$, the actual velocity of the flow, $v$, will be given in terms of $v_q$ through \eqref{vq}, and the charge density will be proportional to $m$ as in \eqref{gothq}. 

Observe that this mapping from neutral to charged solutions implies that two black holes can have the same profile for $m$ even if their charges and rotations are very different. In particular, an almost static (small $v$) but highly charged black hole ($\fq$ slightly below $1/\sqrt{2}$), can have the same shape as a neutral black hole with large velocity if the two solutions have the same value of $v_q$.

\section{Charged rotating black holes: Kerr-Newman at $D\to\infty$}\label{sec:KN}

The solutions of \eqref{mastercharge} for stationary axially symmetric charged lumps on a 2-brane that we construct in this section correspond to the large-$D$ limit of the elusive higher-dimensional generalization of the Kerr-Newman black hole.

\subsection{Solution}

We employ polar coordinates $(r,\phi)$ on the 2-brane. Ref.~\cite{Andrade:2018nsz} found the neutral solution that describes the Myers-Perry rotating black hole, with area-radius profile
\beq\label{RMP2}
\cR(r)=\frac{2}{1+a^2}\lp 1-\frac{r^2}{4}\rp\,,
\eeq
and angular velocity
\beq\label{Oma}
v^\phi=\Omega=\frac{a}{1+a^2}\,. 
\eeq
When we fix the overall scale, \eg by fixing the mass, this is a one-parameter family of solutions with the rotation parameter $a$ varying in $[0,\infty)$. We restrict to non-negative angular velocities without loss of generality. Observe that $\Omega$ varies between $0$ and $1/2$, the latter maximum being reached when $a=1$.

Applying the procedure described at the end of the previous section we obtain a charged rotating solution, with
\beq\label{KNsol}
\cR(r)=\frac{2}{1+\aq^2}\lp 1-\frac{r^2}{4}\rp\,.
\eeq
The constant $\aq$ is not the rotation parameter anymore, but can be regarded as characterizing the spread of the gaussian for $m(r)=\exp\lp\cR(r)\rp$. Now we have a family of solutions with two parameters, $\aq$ and $\fq$, whose range is
\beq
0\leq \aq<\infty\,,\qquad 0\leq \fq<\frac1{\sqrt{2}}\,.
\eeq

The angular velocity is determined by \eqref{vq} as
\beq\label{Omq}
\Omega=\left(1-2\mathfrak{q}^2\right)^{1/4}\frac{\aq}{1+\aq^2}\,,
\eeq
and the angular momentum, charge and horizon entropy for a given mass are
\beq\label{JM}
J=2 \aq \left(1-2\fq^2\right)^{1/4}M\,,
\eeq
\beq\label{QM}
Q=\fq M\,,
\eeq
\beq\label{AM}
S=2\pi \lp 1+\sqrt{1-2\fq^2}\rp M\,.
\eeq
In addition, the temperature and electric potential are\footnote{The dimensionally correct area is actually not proportional to $M$ but to $M^{1+1/(n+2)}$, and $\Omega$ and $T$ are proportional to $ M^{-1/(n+2)}$, but we neglect these differences in the limit $n\to\infty$. Moreover, at large $D$ the actual physical values of $J/M$ and $A_H/M$ are $1/D$ times those in \eqref{JM} and \eqref{AM}, and $T$ is $D$ times \eqref{TM}; here they have all been rescaled to render them finite \cite{Emparan:2016sjk}. Finally, we use units where $16\pi G=1$.}
\beq\label{TM}
T=\frac{\sqrt{1-2\fq^2}}{2\pi  \lp 1+\sqrt{1-2\fq^2}\rp}\,,
\eeq
and
\beq\label{PhiM}
\Phi=\frac{2\fq}{1+\sqrt{1-2\fq^2}}\,.
\eeq
These satisfy
\beq\label{MSQ}
M=TS+\Phi Q\,.
\eeq
Observe that the rotation term $\Omega J$ does not appear in this relation: at large $D$ it  only enters at next to leading order \cite{Emparan:2013moa} since the velocities and momenta along the effective membrane are $\ord{1/\sqrt{D}}$. Since \eqref{QM}--\eqref{PhiM} are fixed by the properties of the charged membrane and therefore are common to all stationary solutions of the effective equations, in order to distinguish different phases we need to consider their rotational properties.

In appendix \ref{app:RNT} we verify that the solution with $\aq=0$, $\Omega=0$ corresponds to the large-$D$ limit of the static Reissner-Nordstr{\"o}m black hole.

\subsection{Uniqueness}

An immediate consequence of our construction is the uniqueness of the solutions: given $J$ and $Q$ for fixed mass $M$, the parameters $\aq$ and $\fq$ are uniquely determined and therefore so is the solution too.

This result is perhaps not unexpected, but given the previous lack of knowledge about these black holes when neither $J$ nor $Q$ are infinitesimally small, it was not obviously foreordained. The uniqueness only holds, though, within the Kerr-Newman class: it will be violated by the charged black bars of the next section (but only at $D\to\infty$) and by other classes of charged black holes, such as charged black rings and bumpy black holes (in all $D\geq 6$).

\subsection{Effects of charge and rotation}\label{sec:effQJ}

Let us now use \eqref{QM} to rewrite \eqref{JM} as
\beq\label{spreadJQ}
\aq=\frac{J}{2M}\frac1{\lp 1-2\lp\frac{Q}{M}\rp^2\rp^{1/4}}\,.
\eeq
This equation serves to illustrate the effect that angular momentum and charge have on the shape of a black hole of a given mass. Increasing the spin $J$ results in a proportionately larger spread of the black hole $\aq$, as is already familiar for rotating black holes in any $D\geq 6$ \cite{Emparan:2003sy}. If we then add charge, we see that the rotational spreading is enhanced. This effect, which, again, we expect to happen in every $D\geq 6$, is naturally attributed to electrostatic repulsion: intuitively, the horizon becomes less gravitationally tight.

We can also see in \eqref{AM} how the charge reduces the horizon area for a given black hole mass, which is a generic phenomenon in all $D\geq 4$. Observe, however, that the presence of rotation does not change the horizon area. As discussed above, this is a leading large-$D$ effect, which can be explicitly observed for Myers-Perry black holes. Relatedly, note that the extremal limit (which, as we said, strictly lies outside the scope of our analysis) depends only on $Q/M$ but not on $J/M$. This is in contrast to the properties of the four-dimensional Kerr-Newman solution, but on the other hand is in consonance with the absence of an extremal rotating limit for singly-spinning black holes in $D\geq 6$.

Consider now solutions with fixed charge-to-mass ratio $\fq$. Eq.~\eqref{Omq} implies that the maximum rotation velocity in this case is
\beq\label{Omegamax}
\Omega_{\textrm{max}}=\frac12 \left(1-2\mathfrak{q}^2\right)^{1/4}\,.
\eeq
For any other value of the rotation, $\Omega$ in \eqref{Omq} is a two-valued function of $\aq$, so there are two possible black holes with the same charge and angular velocity: a `plump' one with $\aq<1$ and a corresponding `pancaked' one with $\aq>1$. In particular, close to the extremal charge limit, for any given small angular velocity we can find two distinct black hole solutions: an almost round one with $\aq\ll 1$, which is very close to the extremal Reissner-Nordstr{\"o}m black hole, and a highly pancaked one with $\aq\gg 1$. Note, however, that these two black holes have very different spins, since the angular momentum $J$ for fixed charge grows monotonically as the gaussian profile broadens with increasing $\aq$.

Eq.~\eqref{spreadJQ} shows that it is possible to have highly pancaked black holes (with large $\aq$) whose angular momentum is small if the charge is sufficiently close (but still not equal) to the maximum value, namely $1/\sqrt{2}-\fq\ll 2\sqrt{2}/a_q^{4}$.
The existence of these black holes with near-extremal charge and small spin whose horizons are pancaked along the rotation plane was first identified in \cite{Caldarelli:2010xz} in any dimension $D\geq 6$. Their extremal limit corresponds to singular solutions of disks of extremal charged dust. In contrast to the method used in \cite{Caldarelli:2010xz}, which only works in the highly-pancaked limit (charged or not), our construction allows to cover the entire phase space of charges (above extremality) and rotations, and thus interpolate continuouly between plump and pancaked solutions.

\section{Charged rotating black bars}\label{sec:bbars}

We can similarly apply the procedure described in sec.~\ref{subsec:statconf} to the neutral black bar solution of \cite{Andrade:2018nsz}, and thereby generate charged rotating black bars, with area-radius profile
\beq\label{Rbar}
\cR(t,r,\phi)=1-\frac{r^2}{4}\lp 1+\sqrt{1-\frac{4\Omega^2}{\sqrt{1-2\fq^2}}}\,\cos\lp 2(\phi-\Omega t)\rp\rp\,.
\eeq
If we employ corotating coordinates
\beq
x_t=r\cos(\phi -\Omega t)\,,\qquad y_t= r\sin(\phi -\Omega t)\,,\label{cartcorot}
\eeq
then the solution \eqref{Rbar} reads
\beq
\cR(x_t,y_t)=1-\frac{x_t^2}{ 2\ell_\perp^2}-\frac{y_t^2}{2\ell_\|^2}\,,
\eeq
where the lengths parallel to the bar and transverse to it are
\beqa
\ell_\|^2&=& \frac{2}{1-\sqrt{1-\frac{4\Omega^2}{\sqrt{1-2\fq^2}}}}\,,\nn\\
\ell_\perp^2&=& \frac{2}{1+\sqrt{1-\frac{4\Omega^2}{\sqrt{1-2\fq^2}}}}\,.
\eeqa
The effect of adding charge to a bar of a given mass and angular velocity is to reduce its length $\ell_\|$ and increase its thickness $\ell_\perp$.

\subsection{Physical properties and non-uniqueness}\label{sec:nuniq}

For these solutions the angular momentum is
\beq
J=\frac{\sqrt{1-2\fq^2}}{\Omega}\,M\,,
\eeq
while the charge, entropy, temperature and potential are given by the same expressions as \eqref{QM}--\eqref{PhiM}.

The angular velocity $\Omega$ varies between $0$ and a maximum $\Omega_{\textrm{max}}$ which is the same as in \eqref{Omegamax}. Therefore the angular momentum of the black bars is bounded below, satisfying
\beq\label{uniqbound}
\frac{J}{M}\geq 2 \lp 1-2\lp\frac{Q}{M}\rp^2\rp^{1/4}\,.
\eeq
This is the region marked in red in fig.~\ref{fig:Phases}.

Unlike the solutions of the previous section, there is only one black bar for given $\Omega$ and $Q/M$. The limit $\Omega\to 0$ yields an infinite, static charged black string. As in the neutral case, long charged bars behave like rigidly rotating solids with $J\sim M \ell_\|^2\Omega$.

When $\Omega=\Omega_{\textrm{max}}$ the solution becomes axisymmetric, with $\ell_\|=\ell_\perp$; actually, we recover the same maximally-rotating charged black hole of sec.~\ref{sec:KN} with $\aq=1$. Thus this solution sits at a bifurcation point in solution space---actually it is a line of bifurcation points, parametrized by $\fq$. We will see later that this family is indeed marked by the appearance of a zero mode at the threshold of a bar-mode instability of the axisymmetric charged black holes.

For $M$, $Q$ and $J$ that satisfy \eqref{uniqbound} we can always find a Kerr-Newman black hole and a black bar with the same values of these conserved charges.
In this range, therefore, black hole uniqueness does not hold. These solutions also have the same entropy (to leading order at large $D$), but even if they are thermodynamically equally preferred, they can differ in their dynamical stability. Indeed, we will see that near the saturation of the bound \eqref{uniqbound} the Kerr-Newman black hole is linearly unstable while the black bar is (most likely) stable.

\subsection{Radiation from charged black bars}

In any finite number of dimensions, a charged rotating black bar will radiate both electromagnetic and gravitational waves. It is easy to estimate that the radiating power into each channel is
\beqa
P^{\textrm{em}}&\sim& G Q^2\ell_\|^2\,\Omega^D\,,\\
P^{\textrm{gr}}&\sim& G M^2\ell_\|^4\,\Omega^{D+2}\,,
\eeqa
where, note, we are measuring the charge in geometric units, hence the factor $G$ in $P^{\textrm{em}}$.

Since we always have $\Omega\leq 1/2$, this radiation at large $D$ is exponentially small, $\sim e^{-D}$, and thus invisible in the perturbative $1/D$ expansion. This is why in our approach we can find black bars as stationary solutions: their decay time is exponentially long in $D$.

For a long black bar, with
\beq
\ell_\|^2\sim \frac{\sqrt{1-2\fq^2}}{\Omega^2}
\eeq
the ratio of radiation into each channel is
\beq\label{ratioemgr}
\frac{P^{\textrm{em}}}{P^{\textrm{gr}}}\sim \lp\frac{Q}{M}\rp^2\frac1{\sqrt{1-2(Q/M)^2}}\,.
\eeq
There are two different factors here that, as charge is added to the bar, enhance the power into electromagnetic radiation relative to gravitational radiation. The overall factor $(Q/M)^2$ accounts for the larger amount of charge that the bar carries. The second factor in \eqref{ratioemgr} is due to the electromagnetic-dipolar vs.\ gravitational-quadrupolar nature of the emission of radiation: as we mentioned above, for a given mass and rotation velocity the bar gets shorter and fatter as charge is added, which reduces the quadrupole moment by a larger factor than the dipole moment.

\section{Quasinormal modes}\label{sec:qnms}

\subsection{Co-rotating zero modes}
\label{sec:corot}
We begin by studying co-rotating zero-mode perturbations, which keep the solution stationary. In this case we can directly adapt the results obtained in \cite{Andrade:2018nsz} for the neutral case. Note that stationarity of the charged perturbations implies that there is no charge diffusion mode.

\subsubsection{Kerr-Newman black hole}

By a direct map from the neutral case, we obtain the co-rotating, zero-mode perturbations in the form
\beq
\cR(r)=\frac{2}{1+\aq^2}\lp 1-\frac{r^2}{4}\rp+\epsilon\,\delta \cR(r) e^{i\nh(\phi-\Omega t)}\,,
\eeq
where
\beq\label{cRLaguerre}
\delta\cR(r)= r^{|\nh|} L_k^{|\nh|}\lp \frac{r^2}{2(1+\aq^2)}\rp\,. 
\eeq
These modes only exist if the width parameter $\aq$ takes the values
\beq\label{azero}
a_{q,c}^2=\ell-1\,,
\eeq
with
\beq\label{elldef}
\ell = 2 k + |m_\phi|\,,\qquad k=0,1,2,\dots\,,\qquad |\nh|=0,1,2,\dots
\eeq
which in turn implies via \eqref{Omq} a quantization condition for the angular velocities at which zero modes appear
\beq
\Omega=\frac{\sqrt{\ell-1}}{\ell}\left(1-2\mathfrak{q}^2\right)^{1/4}\,,
\eeq
(only $\ell\geq 2$ is meaningful) and also for the angular momenta
\beq\label{JMzero}
\frac{J}{M}=2\sqrt{\ell-1} \lp 1-2\lp\frac{Q}{M}\rp^2\rp^{1/4}\,.
\eeq
Observe that the appearance of these modes---which, we will see, mark the thresholds of instabilities---depends on the width of the profile, $\aq$, and not on the charge and rotation separately. In particular this implies that the addition of charge reduces the range where a solution of given rotation is stable. Equivalently, a mode of given $k$ and $|\nh|$ will appear at lower rotation the larger its charge is. Again, these are manifestations of the repulsive effect of charge.

The axisymmetric modes with $\nh=0$ are expected to lead to branches of `bumpy charged black holes'. The fundamental non-axisymmetric modes, with $k=0$ and $\ell=|\nh|\geq 2$ are multipole-bar-mode deformations\footnote{The mode with $k=0$ and $|\nh|=1$ is gauge.}; when $\ell=|\nh|=2$ we recover the linearization of the charged black bar \eqref{Rbar} solution near the maximal angular velocity \eqref{Omegamax}. Indeed  \eqref{JMzero} for $\ell=2$ correctly reproduces the bound on the existence of black bars.

\subsubsection{Black Bars}
The same analysis applies for charged black bars, which have zero modes for 
\beq
\frac{\Omega}{\left(1-2\mathfrak{q}^2\right)^{1/4}}=\frac{\sqrt{2}}{3}\,,\frac{\sqrt{3}}{4}\,,\frac{2}{5}\,\dots
\eeq
or equivalently, for
\beq\label{JMzerobar}
\frac{J}{M}=\frac{n+1}{\sqrt{n}}\lp 1-2\lp\frac{Q}{M}\rp^2\rp^{1/4}\,,\qquad n=2\,,3\,,4\dots
\eeq
(the value $n=1$ corresponds again to the bifurcation with KN black holes). As before, adding charge makes the zero modes appear at slower rotation.

These zero modes, like those found in \cite{Andrade:2018nsz}, create ripples along the bar which, in the limit $\Omega\to 0$ where $\ell_\|\to\infty$, become Gregory-Laflamme modes of a black string. Although we have not managed to solve for non-zero modes of black bars, we expect that they are unstable for $\Omega <\frac{\sqrt{2}}{3} \left(1-2\mathfrak{q}^2\right)^{1/4}$, \ie\ for $J/M>(3/\sqrt{2})\lp 1-2(Q/M)^2\rp^{1/4}$. This is indicated in fig.~\ref{fig:Phases}.

\subsection{Quasinormal modes of the $D\to \infty$ Kerr-Newman black hole}
 
We now turn to the study of the quasinormal modes of the full 
time-dependent equations \eqref{mass continuity}-\eqref{chj}.
We consider linear perturbations of the form 
\begin{align}
	m &= \bar m(r) +  e^{- i \omega t + \nh \phi }\, \delta m(r)\,, \\
	p_r &= \bar p_r(r) +  e^{- i \omega t + \nh \phi }\, \delta p_r (r)\,, \\
	p_\phi &= \bar p_\phi (r)  +  e^{- i \omega t + \nh \phi }\, \delta p_\phi (r)\,,  \\
	q &= \bar q (r)  +  e^{- i \omega t + \nh \phi } \,\fq\,  \delta q(r)\,,
\end{align}
where $\bar m$, $\bar p_r$, $\bar p_\phi$, $\bar q$ 
correspond to the charged, rotating black hole background in sec.~\ref{sec:KN}, concretely, 
\beq
\bar m (r)   = m_0 \exp \left(\frac{2}{1+\aq^2} \lp 1 - \frac{r^2}{4}\rp \right)\, ,
\eeq
and
\beq
\bar p_r(r) =  \partial_r \bar m , \quad 
\bar p_\phi(r) = \Omega \bar m r^2, \quad 	\bar q (r)  = \fq \bar m\,,
\eeq
with $\Omega$ the angular velocity in \eqref{Omq} and $m_0$ an arbitrary constant.

\subsubsection{Charge diffusion perturbations}

It is easy to show that the effects due to charge diffusion decouple from the dynamics. In order to see this, we 
introduce the variable
\begin{equation}
	\delta Q(r) \equiv \frac{1}{\bar m(r)}(\delta q(r)  - \delta m (r)).
\end{equation}
Combining the mass and charge continuity equations, we can easily derive the following 
decoupled equation for $\delta Q$
\begin{equation}
	\delta Q'' + \left( \frac{1}{r} - \frac{r}{1+\aq^2} \right) \delta Q' + 
	\left(i (\omega- m_\phi \Omega) - \frac{m_\phi^2}{r^2} \right) \delta Q = 0 , 
\end{equation}
This the same confluent hypergeometric equation that was found in \cite{Andrade:2018nsz} as governing the spheroidal harmonics at small polar angles. Its regular solutions can be written as 
\begin{align}
%
%
				\delta Q = r^{|\nh|} L_{k-1}^{|\nh|} \left( \frac{r^2}{2 (1+\aq^2)}  \right)
\end{align}
where $k \geq 1$ is an integer related to $\omega$ as
\begin{equation}
\label{w charge diffusion}
	\omega = |m_\phi| \Omega  - i \frac{2 (k-1) + |m_\phi| }{1+\aq^2}\,.
\end{equation}
Solutions with $k=1$, $m_\phi =  0$ are pure gauge.
As mentioned before, there are no co-rotating charge diffusion modes. 

The remaining equations are a set of three, coupled, second order ODEs, in which 
$\delta Q$ and $\delta Q'$ appear as sources. Once a solution for $\delta Q$ has 
been inserted, we can solve for the remaining profiles. 
The easiest way to reconstruct the profiles is to write down an appropriate ansatz 
which reduces the ODEs to algebraic equations for some constant coefficients. 
For the axisymmetric case $\nh=0$, we find
\begin{align}
	\delta a &= e^{- \frac{r^2}{2(1+\aq^2)}}  \sum_{i = 0}^{k}  \delta a^{(i)} r^{2 i}\,, \\
	\delta p_r &= e^{- \frac{r^2}{2(1+\aq^2)}} r \sum_{i = 0}^{k}  \delta p_r^{(i)} r^{2 i}\,, \\
	\delta p_\phi &= e^{- \frac{r^2}{2(1+\aq^2)}} r^2 \sum_{i = 0}^{k}  \delta p_\phi^{(i)} r^{2 i}\,.
\end{align}
For non-axisymmetric modes $|\nh| \geq 1$, the profiles are given by 
\begin{align}
	\delta a &= e^{- \frac{r^2}{2(1+\aq^2)}}  r^{|\nh|} \sum_{i = 0}^{k}  \delta a^{(i)} r^{2 i}\,, \\
	\delta p_r &= e^{- \frac{r^2}{2(1+\aq^2)}} r^{|\nh|-1} \sum_{i = 0}^{k+1}  \delta p_r^{(i)} r^{2 i}\,, \\
	\delta p_\phi &= e^{- \frac{r^2}{2(1+\aq^2)}} r^{|\nh|}  \sum_{i = 0}^{k+1}  \delta p_\phi^{(i)} r^{2 i}\,.
\end{align}

\subsubsection{Gravitational perturbations}

Since charge diffusion perturbations decouple henceforth we consistently set $ \delta q = \delta m $ in order to study perturbations in the mass density and the velocity, which describe the truly gravitational degrees of freedom. Their structure closely
resembles the neutral fluctuations discussed in \cite{Andrade:2018nsz}. In particular, they can also be fully decoupled by means of a 
sixth order operator which factorizes in terms of confluent hypergeometric operators. More concretely, 
we can derive a sixth order equation for $\delta m$ of the form
\begin{equation}
	{\cal L}_1 {\cal L}_2 {\cal L}_3 \delta {\cal R}(r) = 0\,,
\end{equation}
where
\begin{equation}
	\delta {\cal R}(r) =  \frac{ \delta m(r) }{ \bar m(r) }\,,
\end{equation}
and the ${\cal L}_i$ are three commuting differential operators of the same confluent hypergeometric type as in \cite{Andrade:2018nsz},
\begin{equation}
\label{op Li}
	{\cal L}_i = \frac{d^2}{dr^2} + \left( \frac{1}{r} - \frac{r}{1+\aq^2} \right) \frac{d}{dr} + \left( \frac{2 k_i + \nh }{1+\aq^2} - \frac{\nh^2}{r^2} \right)\,.
\end{equation}
The constants $k_i$ are solutions of the cubic equation
\begin{equation}
\label{cubic w}
	\omega^3 + \frac{i \omega^2}{2(1+\aq^2)} P_2+ \frac{\omega}{2(1+\aq^2)^2} P_1+ \frac{i}{2(1+\aq^2)^3} P_0 =0
\end{equation}
where
\begin{align}
	P_2 &= (\beta^2 +1) (3 \ell-4)+6 i \aq \beta m_{\phi } \\
\nonumber
	P_1 &= 2 \aq^2 \beta^2  \left(\ell+3 m_{\phi }^2-4\right)-6 i \aq \beta (\beta^2 +1) (\ell-1) m_{\phi }\,, \\
	& -(\ell-1) \left((\beta^2  (\beta^2 +4)+1) \ell-2 (\beta^2 +1)^2\right)\,,  \\
\nonumber
	P_0 &=  \aq^2 \beta^2  (\beta^2 +1) \left((2-3 \ell) m_{\phi }^2-(\ell-2) \ell\right)-2 i \aq^3 \beta ^{3} m_{\phi } \left(\ell+m_{\phi }^2-2\right) \\
	& \beta^2  (\beta^2 +1) (\ell-2) (\ell-1) \ell+i \aq \beta (\ell-1) \left(\beta ^4 (\ell-2)+4 \beta^2  \ell+\ell-2\right) m_{\phi }\,,
\end{align}
with $\ell$ as in \eqref{elldef}
and we abbreviate
\beq
\beta = (1- 2 \fq ^2)^{1/4}\,.
\eeq
Note that this equation is a cubic in $\omega$ and also in $\ell$ (or $k$).
Since the operators ${\cal L}_i$ commute, all solutions of this cubic are equivalent and can be analized separately. 
Moreover, solving the radial equations we learn that $k$ must be non-negative integers. Thus, \eqref{cubic w} becomes 
a quantization condition for the frequencies, in such a way that, for a given value of $k$ and $m_\phi$, and the black hole parameters,
\eqref{cubic w} determines three possible frequencies associated to them.\footnote{If we keep the degree of freedom $\delta q$ in the analysis, we can derive an eighth-order 
equation factorized into operators of the form \eqref{op Li}, with a quartic equation for $\omega$
which factorizes into \eqref{cubic w} and a linear piece equal to \eqref{w charge diffusion}.}

Eq.~\eqref{cubic w} reproduces our previous result in 
\cite{Andrade:2018nsz} in the neutral case $\beta = 1$. Moreover, in the static limit $\aq=0$ with nonzero charge we can solve to find
\beqa
	\omega &=& - \frac{i}{2}(\ell-2)(\beta^2+1),\\
\omega &=& - \frac{i}{2} (\ell-1)(\beta^2+1) \pm 
	\frac{1}{2} \left[ (\ell-1)[ (\beta^2+1)^2 - \ell (\beta^2-1)^2 ] \right]^{1/2}\,.
\label{statqnm}\eeqa
These frequencies match the results obtained in \cite{Tanabe:2016opw,Bhattacharyya:2015fdk} (provided we account for the different definitions of $\ell$ for vectors and scalars), which used a different approach for the perturbations of static charged black holes at $D\to\infty$. These modes are all stable, but they illustrate a continuing theme in our analysis: the addition of charge reduces the stability of the solution, in this case by decreasing the damping rate of the perturbation. 

The analysis of the solutions to the quantization condition \eqref{cubic w} and the reconstruction of the full 
profiles proceeds in close parallel to \cite{Andrade:2018nsz}. 
In the remainder of this section we record such expressions and highlight the presence of some unstable 
modes.

\subsubsection*{Axisymmetric modes: $\nh=0$}

There are no regular solutions for $\ell = 0$. For $\ell = 2,4,6, \ldots$, all solutions are regular and have profiles given by 
\begin{align}
	\delta p_r &= e^{- \frac{r^2}{2(1+\aq^2)}} r \sum_{i = 0}^{k}  \delta p_r^{(i)} r^{2 i}\,, \\
	\delta p_\phi &= e^{- \frac{r^2}{2(1+\aq^2)}} r^{2} \sum_{i = 0}^{k}  \delta p_\phi^{(i)} r^{2 i}\,.
\end{align}

\subsubsection*{Non-axisymmetric modes: $|\nh| \geq 1$}

\begin{itemize}

\item 
For $k=0$, $\ell = |\nh|$ we find
\begin{align}
	\omega_0 &= \frac{\aq \beta (|\nh|+4) - i (\beta^2+1)( |\nh|-2 ) }{2(1+\aq^2)}	\,,\\
	\omega_\pm &= \frac{(|\nh|-1) [ 2 \aq \beta -i (\beta^2 +1)]}{2(1+\aq^2)} \\ 
	&\quad\pm 	\frac{i}{2(1+\aq^2)} \left \{  (|\nh|-1) (\beta^2 -1)^2 |\nh| +\left(2 \aq \beta -i (\beta^2 +1)\right)^2  \right \}^{1/2} \,.
\end{align}

Modes with frequency $\omega_0$ are regular for $|\nh| > 2$, so they are stable. Modes with 
$\omega= \omega_\pm$ are regular if $|\nh| \geq 1$. All these solutions have profiles 
of the form
\begin{align}
	\delta p_r &= e^{- \frac{r^2}{2(1+\aq^2)}} r^{|\nh|+1} \delta p_r^{(1)}\,, \\
	\delta p_\phi &= e^{- \frac{r^2}{2(1+\aq^2)}} r^{|\nh|+2} \delta p_\phi^{(1)} \,.
\end{align}

\item 
For $k \geq 1$, all solutions are regular and have profiles given by 
\begin{align}
\delta p_r &= e^{- \frac{r^2}{2(1+\aq^2)}} r^{|\nh|-1} \sum_{i = 0}^{k+1}  \delta p_r^{(i)} r^{2 i}\,, \\
	\delta p_\phi &= e^{- \frac{r^2}{2(1+\aq^2)}} r^{|\nh|} \sum_{i = 0}^{k+1}  \delta p_\phi^{(i)} r^{2 i}\,.
\end{align}

\subsubsection*{Near-critical unstable modes}

The zero modes that signal the onset of the instability, already found in \eqref{azero}, occur at
\begin{equation}
	\aq = a_{q,c} \equiv \sqrt{\ell-1}, \qquad \omega = \omega_c =  |\nh|  \Omega   \,.
\end{equation}
In order to examine the solutions near this point, we perform a perturbative analysis letting
\begin{equation}
	\aq = a_{q,c} + \delta \aq, \qquad \omega = \omega_c + \delta \omega\,.
\end{equation}
We find
\begin{align}
	{\rm Re} \, \delta \omega & = 	\delta \aq \frac{4 \beta ^{3} |\nh| \left(-4 (\beta^2 -1)^2-(\beta^2 +1)^2 \ell^2+4 (\beta^2 -1)^2 \ell\right)}{\ell \left((\ell-1) \left((\beta^2 +1)^2 \ell-2 (\beta^2 -1)^2\right)^2+4 \beta^2  (\beta^2 +1)^2 |\nh|^2\right)}\,, \\
	{\rm Im} \, \delta \omega & = \delta \aq \frac{2 \beta^2  (\beta^2 +1) \sqrt{\ell-1} \left\{ (\ell-2) \ell \left((\beta^2 +1)^2 \ell-2 (\beta^2 -1)^2\right)+8 \beta^2  
	|\nh|^2\right \} }{\ell \left((\ell-1) \left((\beta^2 +1)^2 \ell-2 (\beta^2 -1)^2\right)^2+4 \beta^2  (\beta^2 +1)^2 |\nh|^2\right)}\,.
\end{align}
The imaginary part of the perturbed frequency is manifestly positive for $\delta \aq >0$, which shows that 
there are unstable modes for $\aq > a_{q,c}$. For $|\nh| =0$, the real part of the frequency vanishes, so these modes
are static. On the other hand, for $|\nh| >0$ these modes have a non-zero real part so they correspond to bar modes. 
Moreover, we see that for $|\nh| >0$ , ${\rm Re} \, \delta \omega < 0$: the unstable modes rotate in the same sense but more slowly than the black hole.

As the charge increases and $\beta$ decreases from 1 to zero, the value of ${\rm Im} \, \delta \omega/|\delta \aq|$ decreases monotonically towards zero. This implies that in the stable side, where $\delta \aq<0$, the damping rate decreases with increasing charge: this is the same effect as mentioned after \eqref{statqnm}. On the other hand, it also implies that the instability for $\delta \aq>0$ develops more slowly when charge is present; the same behavior was observed for charged black branes in \cite{Emparan:2016sjk}. 

\end{itemize}

In fig.~\ref{fig:Phases} we have only indicated the onset of the instability that appears at slowest rotation, namely the mode $k=0$, $\ell=|\nh|=2$. This perturbation is unstable for all Kerr-Newman black holes with $a_q>1$. As $J$ increases past the thresholds \eqref{JMzero} and \eqref{JMzerobar}, further unstable modes appear in the red region of the diagram.

\section{Outlook}\label{sec:outlook}

The results in this article push our knowledge of higher-dimensional charged rotating black holes significantly beyond previous studies. We have obtained the complete phase space of Kerr-Newman black holes (except the strictly extremal solutions) in the limit $D\to\infty$, and investigated in detail their mode stability. We have also constructed solutions for charged rotating black bars that should be long-lived at finite $D$. A running theme in our study has been that the addition of charge tends to diminish the stability of the solutions, either by reducing the range of parameters over which they remain stable, or by decreasing the damping rate of stable perturbations. The shortcoming that one must remain away from the extremal limit (which need not commute with $D\to\infty$) does not seem unsurmountable, but it requires careful separate treatment which we postpone for future work.

Our approach focuses on the geometry close to the rotation axis of the black hole, within polar angles $\theta\approx 1/\sqrt{D}$. The study of this region is enough to obtain the black hole mass, entropy, angular momentum and charge with factorial accuracy $\ord{D^{-D}}$, and to capture the wavefunctions and frequencies of quasinormal modes. Still, it would be interesting to also know the shape of the horizons of these black holes at all polar angles down to $\theta=\pi/2$. Ref.~\cite{Mandlik:2018wnw} obtained exact solutions of the membrane equations of \cite{Bhattacharyya:2015fdk} with either infinitesimal charge or infinitesimal rotation. The configurations with both finite charge and finite rotation may require numerical solution of the effective equations. 

We expect the existence of other black holes with charge and rotation: charged versions of the `bumpy' black holes of \cite{Emparan:2003sy,Dias:2009iu,Dias:2014cia,Emparan:2014pra} (indeed we have obtained in sec.~\ref{sec:corot} the zero-modes with $k>0$ and $\nh=0$ that create $k$ bumps on a Kerr-Newman black hole); charged black rings in $D\geq 6$ (analyzed in \cite{Caldarelli:2010xz} and in \cite{Chen:2017wpf}); and other solutions that combine the previous ones in configurations with disconnected horizons. Further extensions of our analysis also seem feasible, such as the inclusion of charge along with several rotations, possibly in all independent rotation planes in odd $D$ \cite{Tanabe:2016opw}, and the construction of black holes with $p$-form dipoles \cite{Caldarelli:2010xz,Emparan:2010ni,Armas:2013aka}. We expect that the large-$D$ expansion will be useful for charting the space of these solutions and their stability.

\section*{Acknowledgements}
This work is supported by ERC Advanced Grant GravBHs-692951 and MEC grant FPA2016-76005-C2-2-P.


\appendix

\section{Reissner-Nordstr{\"o}m-Tangherlini solution as a gaussian membrane}\label{app:RNT}

Let us consider the Reissner-Nordstr{\"o}m-Tangherlini solution in $D=n+5$ dimensions \cite{Tangherlini:1963bw}
\beq\label{rnt}
ds^2=-f(\hat{r}) dt^2 +f(\hat{r})^{-1}d\hat{r}^2+\hat{r}^2d\Omega_{n+3}^2
\eeq
with
\begin{align}
f(\hat{r})=1-\frac{\mu}{\hat{r}^{n+2}}+\frac{q^2}{2\hat{r}^{2n+4}}\,,
\end{align}
and
\beq
A_t=-\frac{q}{\hat{r}^{n+2}}\,.
\eeq
We set units such that $\mu=1$ and introduce the parameter $\mathfrak{q} \in [0,1/\sqrt{2}]$ as in \eqref{gothq}. We write the sphere $S^{n+3}$ as a fibration of $S^{n+1}$ over a disk,
\begin{align}
ds^2=-f(\hr) d\htt^2+\frac{d\hr^2}{f(\hr)}+\hr^2 d\theta^2+\hr^2\sin^2\theta d\hph^2+\hr^2\cos^2\theta d\Omega_{n+1}\,,
\end{align}
and  introduce
\beq
\rho=\hat{r}\cos\theta\,,\qquad r=\sqrt{n}\hat{r}\sin \theta\,.
\eeq

Taking $n\rightarrow \infty$ we focus on small polar angles $\theta\approx \ord{1/\sqrt{n}}$. Introducing $\sR=\rho^n$ we have
\beq
\hat{r}^{n}\simeq\sR\,  e^{\frac{r^2}{2}}\,,
\eeq
and expanding the metric \eqref{rnt} in $1/n$ we obtain
\beq
ds^2\simeq -A d\htt^2+\frac1{n^2}\frac{d\sR^2}{A \sR^2}+\frac1n\lp 1+\frac{r^2}{n} \frac{ e^{-r^2/2}}{A\sR}\lp 1-\frac{\fq^2  e^{-r^2/2}}{2\sR}\rp\rp dr^2+\frac{r^2}{n}d\hph^2+\br^2 d\Omega_{n+1}\,, 
\eeq
with
\beq
A=1-\frac{e^{-\frac{r^2}{2}}}{\sR}+\frac{\mathfrak{q}^2\,e^{-r^2}}{2\sR^2}\,.
\eeq
Change now to the Eddington-Finkelstein (null) time $t$,
\beq
t=\htt-\frac1n\int\frac{d\sR}{A\sR}\,,
\eeq
so that
\beq
dt=d\htt-\frac1n\frac{d\sR}{A\sR}-\frac{r}{n} \frac{ e^{-r^2/2}}{A\sR}\lp 1-\frac{\fq^2  e^{-r^2/2}}{2\sR}\rp dr\,,
\eeq
to find that the solution takes the form of \eqref{AFch} with
\beqa
m(r)&=&e^{-r^2/2}\,,\nn\\
q(r)&=&\mathfrak{q}\, e^{-r^2/2}\,,\\
p_r(r)&=&-r\, e^{-r^2/2}\,.\nn
\eeqa
Since we have $p_i=\nabla_i m$, the velocity is zero and therefore this is the same as the static limit of \eqref{KNsol}.

\end{document}